# Chaos in oscillatory heat evolution accompanying the sorption of hydrogen and deuterium in palladium[1]

Erwin Lalik
*Jerzy Haber Institute of Catalysis and Surface Chemistry, Polish Academy of Sciences, ul. Niezapominajek, 30-239 Krakow, Poland;, E-mail: nclalik@cyf-kr.edu.pl,*

Aperiodic oscillations in the sorption of hydrogen or deuterium in metallic palladium powder have been observed, and a novel method to confirm their deterministic rather than random character has been devised. For that purpose, a theorem relating the square of a function, with the derivative and integral with variable upper limit of the same function has been proved and proposed to be used as a base for a chaos-vs-random test. Both the experimental and the computed time series may be tested to detect determinism. The result is a single number within the interval [0,2]. The test is designed in such a way that its result is close to zero for the datasets that are deterministic and smooth, and close to 2 for the datasets that are non deterministic (random) or non smooth (discrete). A large variety of the test results has been obtained for the calorimetric time series recorded in thermokinetic oscillations, periodic and quasiperiodic, accompanying the sorption of $H_2$ or $D_2$ with Pd as well as for several non oscillatory calorimetric curves recorded in this reaction. These experimental datasets, all coming form presumably deterministic processes, yielded the results clustering around 0.001. On the other hand, certain databases that were presumably random or non smooth yielded the test results from 0.7 to 1.9. Against these benchmarks, the examined, experimental, aperiodic oscillations gave the test results between 0.004 and 0.01, which appear to be much closer to the deterministic behavior than to randomness. Consequently, it has been concluded that the examined cases of aperiodic oscillations in the heat evolution accompanying the sorption of $H_2$ or $D_2$ in palladium may represent an occurrence of mathematical chaos in the behavior of this system. Further applicability and limitations of the applied test have also been discussed, including its intrinsic inability to detect determinism in the discrete time series.

**The recent finding of periodic and quasiperiodic oscillations in the rate of heat evolution (thermokinetic oscillations) accompanying the sorption of gaseous $H_2$ or $D_2$ in metallic palladium, made one expect a chaotic dynamics as well to occur in this reaction. Indeed, in several experiments, the aperiodic oscillations have been recorded under specified conditions in this system, however, their chaotic, i.e., deterministic, character needed to be confirmed mathematically. A theorem has been formulated and proved, and then used as a base for a novel random–vs-chaos testing procedure. The results of this test for the aperiodic time series in question have been compared to the results of the same test performed for a large body of deterministic, the periodic and quasiperiodic, thermokinetic time series, previously recorded in both the $H_2$/Pd and the $D_2$/Pd system. For a further comparison, certain random databases have also been subjected to the same testing procedure. The results for the examined aperiodic time series turned out to be much closer to the deterministic oscillations than to the random datasets, indicating a deterministic character of the former. This seems to confirm the occurrence of chaotic oscillations of heat evolution in the reaction of $H_2$ and $D_2$ with Pd.**

## I. INTRODUCTION

Oscillatory heat evolution accompanying the sorption of hydrogen or deuterium in metallic palladium has recently been described in two reports.[1,2] The thermokinetic oscillations can been induced by admixing the hydrogen (or deuterium) with ca. 10 % vol. of an inert gas, such as He, Ne, Ar, Kr or $N_2$, prior to its contact with palladium. The frequencies turned out to be functionally dependent on the atomic parameters (the first ionization potential and the square root of atomic mass) of the inert gases being used in the reaction.[2] Apart from specifying conditions for the oscillatory kinetics to occur, the calorimetric experiments also demonstrated a consistent invariance of the value of the heat of sorption of $H_2(D_2)$ in Pd, which remains the same over a whole range of frequencies observed, independent of whether the oscillations are periodic or quasiperiodic, and irrespective of the process duration. Here, we report the occurrence of aperiodic dynamics in oscillatory heat evolution in the H(D)/Pd system and we propose a mathematical test to show that the calorimetric time series in question actually represent mathematical chaos rather than a random noise.

Several mathematical methods have been proposed to be used as a chaos-vs-random test. In the effort to detect the symptoms of chaotic behavior in a given dataset (orbit), these methods differ by applying their algorithms either directly to the time series or to the representations in the phase space. Falling into the first category is the Lyapunov exponent $\lambda$ that measures an extent of the so called sensitive dependence on initial conditions (also known as the butterfly effect) which is a defining characteristic of the mathematical chaos.[3,4] The method of surrogate data creates a set of supplementary datasets by rearranging the original time series in such a way as to retain its linear statistical characteristics (like the mean and variance) in order to be able to compare certain statistics, like the correlation dimension,[5] computed for the original data, to those computed for the surrogate datasets. Basing on the results of such comparison, there can be rejected or accepted a null hypothesis of the original data as being generated by a linear Gaussian (stochastic) process.[6] Among the phase space using methods, the correlation exponent $\nu$ measures the spatial correlation of random points on the trajectory.[5] The method of information dimension $D_I$ begins with a partition of phase space into cells of arbitrary dimension $\varepsilon$ and then applies the Shannon entropy formula to account for the probability of a point to fall within the given cell[7]. In general, therefore, these seemingly most often applied methods make use of certain intrinsic properties of chaos itself rather than of the fundamental fact that chaos is related to a mathematical function whereas randomness, by definition, is not.

In this article, a new mathematical relationship is being outlined and proposed to be used as a simple mathematical test that addresses this fundamental difference between chaos and randomness in certain categories of datasets, both experimental and computed. The test has been applied to the aperiodic time series recorded in the oscillatory sorption of $H_2$ and $D_2$ in palladium. The results have been judged against a large body of results of the same test obtained for periodic and quasiperiodic oscillations in the $H_2(D_2)$/Pd system and turned out close enough to the level indicative of deterministic character. The occurrence of mathematical chaos in the thermokinetic oscillations accompanying the sorption of hydrogen in the metallic palladium has been, therefore, confirmed. To our best knowledge, chaotic dynamics in this system has not been reported to date.

## II. OUTLINE OF CONCEPT

### A. Mathematical description

Let us consider a function $f(x)$ that is smooth, continuous and square integrable within an interval $[a,b]$. Let $f'(x)$ be the first derivative of $f(x)$:

$$f'(x) = \frac{d[f(x)]}{dx} \tag{1}$$

and $g(x)$ be a definite (cumulative) integral of the $f(x)$ within an interval $[a,b]$ with the variable upper limit

$$g(x) = \int_a^x f(t)dt \tag{2}$$

The pointwise product $p(x)$ of these two functions, $f'(x)$ and $g(x)$, can be written as

$$p(x) = (f'g)(x) = f'(x)g(x) \tag{3}$$

The **Theorem 1** then states that if the values of the function $f(x)$ are zero at the points $a$ and $b$ than the following relation holds:

$$\int_a^b p(x)dx = \int_a^a (f'g)(x)dx = -\int_a^b [f(x)]^2 dx \tag{4}$$

In the other words, the areas under the curves $p(x)$ and $[f(x)]^2$, within the interval $[a,b]$, are equal in absolute value but opposite in sign (cf. Scheme 1).

**Proof.** The definite integral of the pointwise product $p(x)$ within the interval $[a,b]$:

$$\int_a^b f'(x)g(x)dx = \int_a^b \left( \frac{d[f(x)]}{dx} \int_a^x f(t)dt \right) dx \tag{5}$$

can be integrated by parts so we obtain

$$\int_a^b \left( \frac{d[f(x)]}{dx} \int_a^x f(t)dt \right) dx = [f(x)g(x)]_a^b - \int_a^b \left( f(x) \frac{d}{dx} \int_a^x f(t)dt \right) dx \tag{6}$$

From the fundamental theorem of calculus we have

$$\frac{d}{dx}\int_a^x f(t)dt = \frac{d}{dx}\int_a^x f(x)dx = f(x) \tag{7}$$

and therefore we can rewrite the expression (6) as follows

$$\int_a^b \left( \frac{d[f(x)]}{dx} \int_a^x f(t)dt \right) dx = [f(x)g(x)]_a^b - \int_a^b [f(x)]^2 dx \tag{8}$$

We remember that $f(a) = 0$ and $f(b) = 0$ and therefore the middle term is also equal to zero

$$[f(x)g(x)]_a^b = f(b)g(b) - f(a)g(a) = 0 \tag{9}$$

which leads us to the desired relation:

$$\int_a^b \left( \frac{d[f(x)]}{dx} \int_a^x f(t)dt \right) dx = -\int_a^b [f(x)]^2 dx \tag{10}$$

**B. The essence of the test**

For brevity, subsequently the letters $P$ and $S$ will be used to denote respectively the left and in the right hand side of the equation (10). It follows from the equation (10), that, since $S$ is always positive (resulting from squared values), then $P$ must always be negative, and so $S = |P|$. The essence of the proposed test consists of using the departure from the latter equality as a measure of deviation from determinism. Since the hypothesis of Theorem 1 assumes that the $f(x)$ is a smooth function rather than a random sequence, so the equation (10) holds for a functional relation but not for a random noise. In practice, it means, the difference between the $S$ and the absolute value of $P$ should approximately be zero for a functional relation (provided that it is continuous and smooth), but it should be non zero for a random data. Both sides of the expression (10) can be easily evaluated numerically for practically any given dataset treated as a function $f(x)$ for the purpose. Once the values of $S$ and $P$ are each calculated with sufficient precision, they can be compared to one another to check for closeness of their sum to zero (remember that $S$ and $P$ have opposite sign).

It is also desirable for the test to be independent of the size and the kind of examined datasets. Therefore, rather than using merely a difference $S - |P|$, the absolute relative difference (ARD), denoted as $D$, can be calculated according to the following formula:

$$D = \left| \frac{S - |P|}{0.5(S + |P|)} \right| \tag{11}$$

The difference of $S$ and $P$ is thus weighted against their arithmetic mean. The formula (11) ensures that $D$ is always positive, as well as that it cannot be larger than 2, since on substituting $S/|P| = a$ we can rewrite (11) as

$$D = 2\left| \frac{a-1}{a+1} \right| \tag{12}$$

and so for $a$ tending to infinity (for $S$ and absolute of $P$ largely different) the $D$ value approaches 2, as we have

$$\lim_{a\to\infty}\left[ 2\left| \frac{a-1}{a+1} \right| \right] = 2 \tag{13}$$

On the other hand, for the datasets meeting the conditions of the Theorem 1, that is, for $S$ very close to absolute of $P$, $a$ will be close to 1, and so

$$\lim_{a\to 1}\left[ 2\left| \frac{a-1}{a+1} \right| \right] = 0 \tag{14}$$

**C. Practical algorithm**

A graphical illustration of the Theorem 1 can be seen in Scheme 1. An arbitrary polynomial, plotted over the interval $[a,b]$, represents a starting function $f(x)$ which is smooth, continuous and square integrable, with $f(a) = f(b) = 0$. The arrows are used to convey sense of a sequence (however arbitrary) of the mathematical operations. Thus the function $f(x)$ is being squared (1), integrated with a variable upper limit (2), and having its first derivative taken (3), yielding new functions, respectively, $[f(x)]^2$, $g(x)$ and $f'(x)$. In the operation (4), the integral $g(x)$ and the derivative $f'(x)$, are used to determine their stepwise product denoted as $p(x)$. The definite integrations from $a$ to $b$ of the $p(x)$ and $[f(x)]^2$ are respectively performed as operations (5) and (6). The resulting areas under the curves of $p(x)$ and $[f(x)]^2$ ($P$ and $S$) are shown in red, and their numerical values may now be compared to one another. For the polynomial used as the illustration in Scheme 1 the sequence of operations (1) to (6) results in $S$ = 1381.209 and $P$ = -1381.085 from which, using the formula (11), we obtain $D$ = 0.0000879.

## III. CHARACTERISTICS OF THE DATASETS TESTED

**A. Scope of the datasets used**

As we could see in the last section, the polynomial of Scheme 1, as perfectly deterministic, smooth and continuous as it is by definition, still yielded the $D$ value of 0.0000879, that is, very close, but not exact zero, no doubt due to a limited precision of numerical computations. For a real life dataset, however, which is never free of a measure of noise, and so even less perfectly deterministic, its score on $D$ may expectedly be higher. The question as to how far from zero a dataset might score and still qualified as deterministic requires a reference set of benchmarks to compare against. To prepare such reference scale, one has to calculate a range of $D$ values, spanning from a minimum $D$ (close to zero), yielded for that purpose by computed and deterministic datasets, to the maximum $D$ values (closer to 2), representing knowingly random or non continuous datasets. It is crucial for such scale also to include experimental, deterministic data in order to provide the appropriate benchmarks. In this work, we use a total of 111 probe datasets, of which 100 are experimental and the other 11 have been computed. The bulk of the experimental data consists of 95 deterministic datasets, i.e., the calorimetric time series obtained for both the Pd/H and the Pd/D system, representing either regular oscillations, periodic or quasiperiodic (cf. Fig. 1b and c), or non-oscillatory (cf. Fig. 1a) calorimetric curves. Both the oscillatory and the non-oscillatory curves are continuous and smooth. They are used to establish the benchmarks for experimental and deterministic $D$ level, that is, a level of $D$ expected for a real-life deterministic result. The remaining part of the experimental datasets consists of a group of 5 visibly aperiodic oscillations observed in both the Pd/H and the Pd/D systems (cf. Fig. 2). It is the main goal of this work to find out whether these 5 aperiodic calorimetric time series are chaotic or random. Another important reference point on the $D$-scale should represent the computed, deterministic chaos (cf. Fig. 3). A measure of $D$ values for deterministic chaos can be assessed using the chaotic solutions to a group of 4 different sets of differential equations obtained numerically (see below). On the higher end of the scale, for the datasets that by definition do not match Theorem 1, the question is how closely should they approach the value of 2 on their $D$-test score (cf. equation 13). To this end, the appropriate $D$ values have been calculated for a group of 4 genuinely random datasets, as well as for a 2 chaotic discrete maps, the latter being deterministic but not continuous (cf. Fig. 4). Below the origins of these datasets are described in more detail.

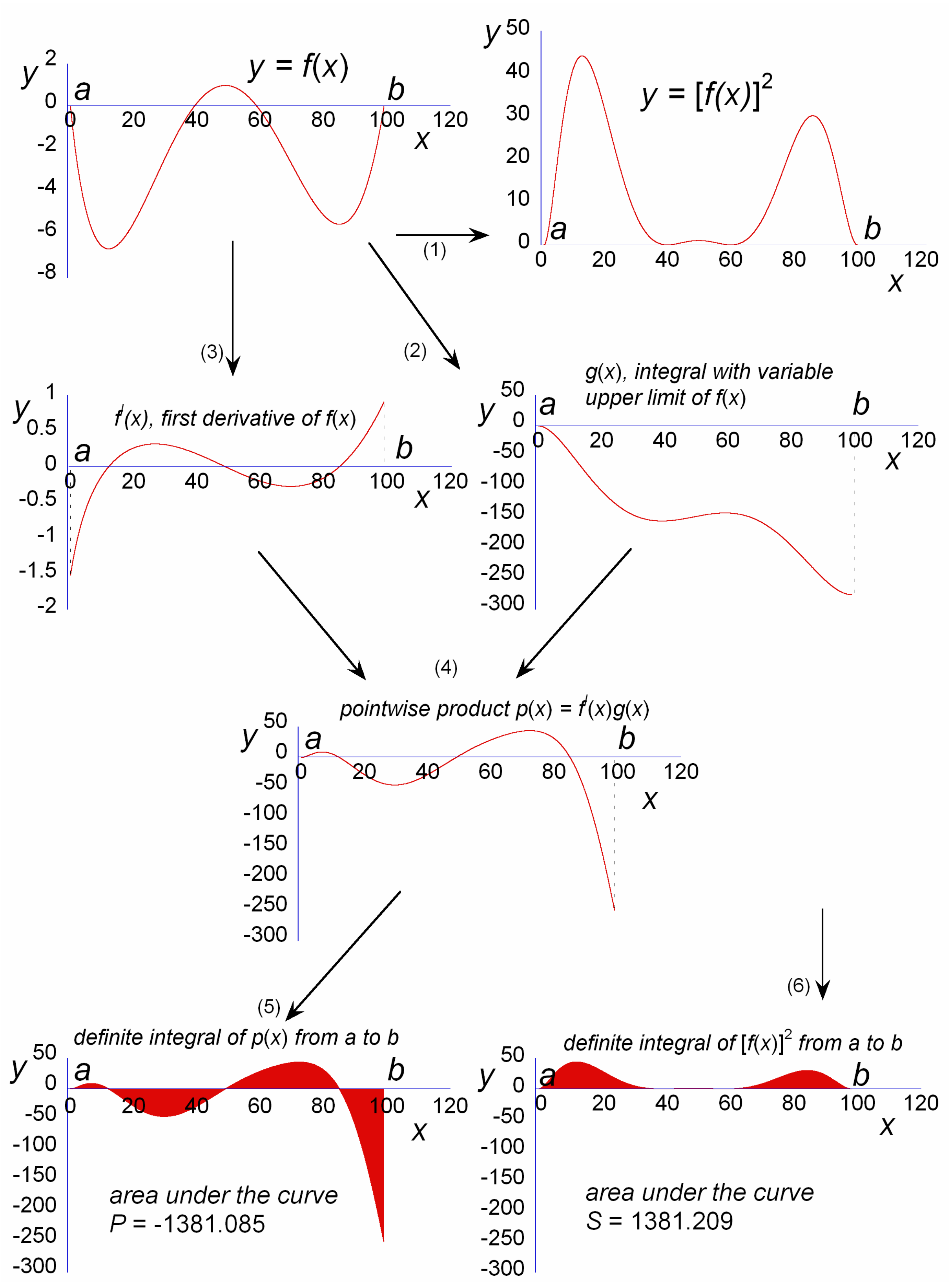

**Scheme 1.** Mathematical operations representing the essence of Theorem 1. Starting from a polynomial $f(x)$ in the upper left, the network eventually leads to a comparison of two definite integrals, that of the square $[f(x)]^2$ and that of the pointwise product $p(x)$, which are equal in absolute values but opposite in signs, according to the theorem. Note, that both the definite integrals (cf. areas in red denoted $P$ and $S$) are represented is the same scale.

**B. Experimental oscillations**

The experimental datasets include calorimetric curves representing the thermokinetic oscillations recently found to appear during the sorption of gaseous $H_2$ or $D_2$ in metallic palladium powder.[1,2] The process is weakly exothermic, and the data represent oscillations of the rate of heat evolution accompanying the reaction, monitored *in situ* with a gas flow-through microcalorimeter. This experimental procedure has been described in detail in Ref. 1. There are several reasons for the thermokinetic oscillations of sorption of hydrogen in Pd to provide the appropriate probe datasets to be used with the formula (10). First, it is natural for the calorimetric curves to start and to end at zero, so the condition of $f(a) = f(b) = 0$ in the hypothesis of Theorem 1 is readily met. Second, the calorimetric curves obtained in the experiments are practically noiseless, in other words they are smooth. Thirdly, the time series may expectedly be continuous, as the heat of reaction is produced continuously, and data are collected every one second. Finally, the process can exhibit all kinds of different oscillatory dynamics and also the microcalorimetric experiments show a very high degree of reproducibility for both the periodic and the quasiperiodic oscillations.[1,2] The examples of the calorimetric curves illustrating the above points are shown in Figs. 1a-c.

Fig. 1a shows the calorimetric curve representing the rate of heat evolution in a typical non-oscillatory sorption experiment in the Pd/$H_2$ system. It can be seen that the process starts and ends at zero, and also that the calorimetric curves obtained with this type of microcalorimeter are practically noiseless. The heat evolution initiates at the moment when hydrogen is being admitted to the flow-through microcalorimetric cell filled with palladium powder. As hydrogen comes into contact with palladium, the calorimetric curve is raising from its initial zero position to a level representing certain rate of heat evolution. It may also mark the onset of oscillations if an admixture of ca. 10 % vol. of inert gas has been added to hydrogen prior to its contact with Pd. In such case, Fig. 1b ($N_2$ admixture) shows that that sustainable periodic oscillations start immediately after the beginning of sorption, and they continue unabated until the end of reaction, signaled by cessation of the heat evolution. At this point the sample of Pd-powder is considered to be saturated with hydrogen, so that, as the sorption is completed, no more heat evolution is being detected, and in most cases the system remains at zero level as long as the same hydrogen/nitrogen gas mixture is flowing through the calorimetric cell. Eventually, switching to the flow of pure nitrogen initiates desorption of hydrogen and, by the same token, triggers an endothermic desorption effect (not shown).

Figure 1c represents the quasiperiodic oscillations at conditions similar to those of Fig. 1b. In fact the two experiments differ only in the way the hydrogen gas flow rate is being initially set up with mass flow controller: a slow adjusting of the knob results in the periodic oscillations to follow, an instant adjustment of the same produce the quasiperiodic dynamics. The quasiperiodic character of oscillations in Fig. 1c can be attested to by the presence of two incommensurate frequencies that can be seen in its power spectrum shown in Fig. 1d (magenta line). For comparison, the black line in Fig 1d shows the power spectrum for the periodic oscillations (that of Fig. 1b), revealing only one incommensurate frequency.

Integration of the calorimetric time series, such as those in Figs. 1a-c, oscillatory or otherwise, yields the total of heat evolved upon sorption of $H_2$ or $D_2$ in the Pd sample. The total heat remains constant, irrespective of whether the oscillations occur or not, and it is also independent of either the dynamic character (periodic o quasiperiodic) or duration of the oscillatory sorption.[1,2] The invariance of the heat of sorption indicates that the thermokinetic oscillations is a kinetic phenomenon that does not alter the thermodynamics of sorption in the Pd/$H_2$($D_2$) system.

Fig. 2 represents the non periodic oscillations. In contrast to the periodic and quasiperiodic experiments, the non periodic time series were obtained during the stages of the Pd sample being saturated with hydrogen, but with the reaction mixture still flowing through the calorimetric cell ($H_2/N_2$ and $D_2$/Kr for Figs. 2a and c respectively). Although in majority of cases, no heat evolution is being detected at the state of saturation, yet it turned out that it is possibly to induce oscillatory heat evolution in this stage by connecting the system to a water aspirator with somehow irregular suction. It appears that the oscillations represent a non periodic sorption-desorption process which apparently can proceed indefinitely. The non periodic character in this case can be confirmed by the very noisy power spectra in

Fig. 2b and c, with poorly resolved features indicative of chaos.[8]

**C. Computed chaotic time series**

These were obtained by numerical integration of several systems of differential equations, including the Lorenz, forced Duffing, forced van der Pol and Rossler equations. A Quickbasic program for Runge-Kutta integration has been taken from Ref. 9 (modified by the present author to enable writing of integration results into a data file). The parameters of these equations were selected so as to ensure for chaotic solutions to be produced on integration (confirmed by power spectra and phase portraits; not shown). The resulting time series showing the *y* - values vs. time are represented in Figs. 3a-d See Appendix A in the supplementary materials[10] for the parameters and initial conditions of the numerical integrations.

**D. Random sequences**

Random generator has been applied to obtain the random Fibonacci (RNF) sequences,[11] of which the initial fragments are shown in Fig. 4a. A simple Quicbasic program for generating the random Fibonacci sequences has been written by the author. Fig. 4b shows a fragment of the "classic" random sequence of $\pi$ digits, that is, the digits after decimal point of the $\pi$ number. The first 20000 digits after decimal point have been used and they were obtained using an application PiW by H. Smith.[12]

**E. Discrete sequences**

The Henon map and the logistic map (Fig. 4c and d respectively) are the results of numerical iterations of the Henon and logistic difference equations (cf. Appendix A in the supplementary materials[10] for parameters of iterations).

## IV. APPLICATION TO EXPERIMENTAL AND COMPUTED TIME SERIES

**A. Dataset preparation**

In order to satisfy the hypothesis of Theorem 1, great care has been taken to ensure, that for all the datasets tested, their starting and ending points were either equal to zero or as close to zero as possible. For most of the datasets that resulted from microcalorimetric experiments no action was necessary. For those datasets, however, that did not naturally started and ended at zero, this condition could be achieved by appropriate truncating at both ends of the dataset. The *D*-testing procedure has a certain degree of tolerance for departure from zero at both ends of the datasets being tested. In fact it is more forgiving for a starting point not being strictly at zero than for the ending point in this respect. In practice, it turned out to be enough for the ending point of a time series to be as close to zero as 0.02.

For the time series that begin and end at zero, but otherwise do not cross the abscissa, as it is the case for all the non-oscillatory, periodic and quasiperiodic time series (cf. Fig. 1a-c), only a single value of *D* has been calculated for each such dataset. However, for the time series repeatedly crossing the abscissa (cf. Figs. 2a and c, 3a-d), the testing procedure has been applied several times, each time producing slightly different values of *D* for the same dataset, depending on which zero-point (abscissa crossing) has been chosen for ending. Effectively, therefore, for each of the aperiodic and of the computed-chaotic dataset, the final *D*-score is actually a mean from a range of *D*-values obtained for successively shortened "versions" of the same dataset.

**B. The test results**

Fig. 5 compares the *D*-scores for all datasets tested, represented in the logarithmic *y*-scale. Different colors are used each to represent a separate category of datasets. (See Table I in Appendix B in the supplementary materials[10] for a list of D-values for all the dataset tested). There is a clear difference between the random *D*-values and the experimental deterministic *D*-values, the former larger by three orders of magnitude than the latter. The *D*-values for the four random and the two discrete sequences (respectively in yellow and brown) are shown in the far right, ranging between 0.7 and 1.9. The highest *D*-values were obtained for the iteration maps, respectively 1.9 and 1.5, for the Henon and the logistic map, whereas the three examples of random Fibonacci sequences yielded $D = 0.7$, 1.0 and 1.2. The classic among the random datasets, a sequence of digits of the number $\pi$ scored on average 0.7 in *D*-value, the average taken over a range $\pi$-digits sequences of various lengths (but always starting and closing at 0).

On the other hand, the majority of the deterministic experimental datasets (from orange to red) tend to cluster around 0.001, albeit with certain outliers reaching as low as 0.00003 (cf. no 45 in Table I[10]), or as high as 0.009 (cf. no 92 in Table I[10]). The average *D*-value for the non-oscillatory databases is 0.00044, while the average *D*-values for the oscillatory experiments are respectively 0.00101 ($H_2/N_2$ periodic), 0.00078 ($H_2/N_2$ quasiperiodic), 0.00153 ($D_2/N_2$ periodic) and 0.00189 ($D_2/Kr$ periodic). The non-oscillatory datasets, therefore, scored lower in the test compared to the periodic and quasiperiodic oscillatory experiments. Among the latter, however, the dynamic character (periodic or quasiperiodic) or the composition of reaction mixture ($H_2/N_2$, $D_2/N_2$, $D_2/Kr$) seem to have little effect on *D*-values obtained.

The aperiodic, experimental datasets (red) range between 0.0040 and 0.0097 (average 0.00673), and thus place themselves close to the upper values obtained for the periodic experiments, although clearly higher than the periodic averages. They are also considerable higher than the computed, chaotic series (turquoise), which scored between 0.0003 and 0.001, with the highest *D*-value of the four obtained for the Lorenz equation. At the same time, however, they are two orders of magnitude lower than the lowest values obtained for the random datasets, which are around 0.7. In fact, visual inspection of Fig. 5 shows that the red bars are clearly much closer to the "deterministic" level on the left than to the "random" one on the right hand side of the figure. It seems therefore reasonable to conclude, that the aperiodic datasets can be qualified as deterministic. Consequently, these datasets may represent the occurrence of mathematical chaos in the oscillatory kinetics of the sorption of hydrogen in palladium.

## V. DISCUSSION

Application of the test based on Theorem 1 seems to confirm the deterministic, chaotic, character of the aperiodic time series representing the thermokinetic oscillations during saturation stage of the sorption of hydrogen in palladium. The values of *D* obtained for the corresponding datasets are relatively close to those obtained for the periodic and quasiperiodic oscillations recorded in both the $H_2$/Pd and the $D_2$/Pd systems. On the other hand, the test expectedly failed to detect determinism in the Henon and logistic maps, since they are neither continuous nor smooth functions. It is therefore more a test for smoothness rather than determinism, however, it can be argued, that smoothness itself implies determinism.[13,14]

The test can be applied to real life datasets, it does not require any surrogate datasets to be created, and it yields a single number as result. The only requirement for the dataset itself is that it has to start and end at zero. The testing procedure is simple and very quick, and any graphing application (the author uses KaleidaGraph) can easily be used for the purpose. Indeed, with a use of a little macro, it takes less than 10 seconds to obtain the final *D*-value.

On the other hand, a meaningful use of the test requires certain prior knowledge of the source of the datasets tested. For the aperiodic oscillations described in this article, the relatively low *D*-value confirms their deterministic, presumably chaotic dynamics. However, this conclusion is supported by a comparison to a vast body of *D*-testing results obtained for periodic and quasiperiodic oscillations in the $H_2$/Pd and $D_2$/Pd systems (cf. Fig. 5). In fact, a low *D*-value was an expectation here, basing on the prior knowledge of oscillatory kinetics of the reaction. But the opposite may also be the case. For instance, when a deterministic, discrete map is expected from the experiment, the actual low score in *D* may possibly indicate that an adverse, but a deterministic in nature process, or a perturbation, is taking place in the experimental system.

## VI. CONCLUSIONS

Chaotic (aperiodic), thermokinetic oscillations have been recorded calorimetrically in the reaction of sorption of hydrogen in palladium. A theorem relating the definite integral of a square of a function $f(x)$ with the definite integral of a pointwise product of the first derivative of $f(x)$ and the cumulative integral (with variable upper limit) of $f(x)$ has been formulated and proved. Basing on the theorem, a procedure to distinguish random from determinism has been proposed. The results of such test applied for the aperiodic, calorimetric time series in question suggest, that they represent the occurrences of chaotic dynamics in the reaction of sorption of hydrogen or deuterium in the metallic palladium. These preliminary results seem to confirm a diagnostic value of the newly proposed test. On the side of limitations, however, its meaningful application requires a considerable prior knowledge and understanding of the experimental system from which the tested datasets originate.

## ACKNOWLEDGMENTS

The author is grateful to Dr A. J. Groszek and to Microscal Ltd., London, for support and assistance.

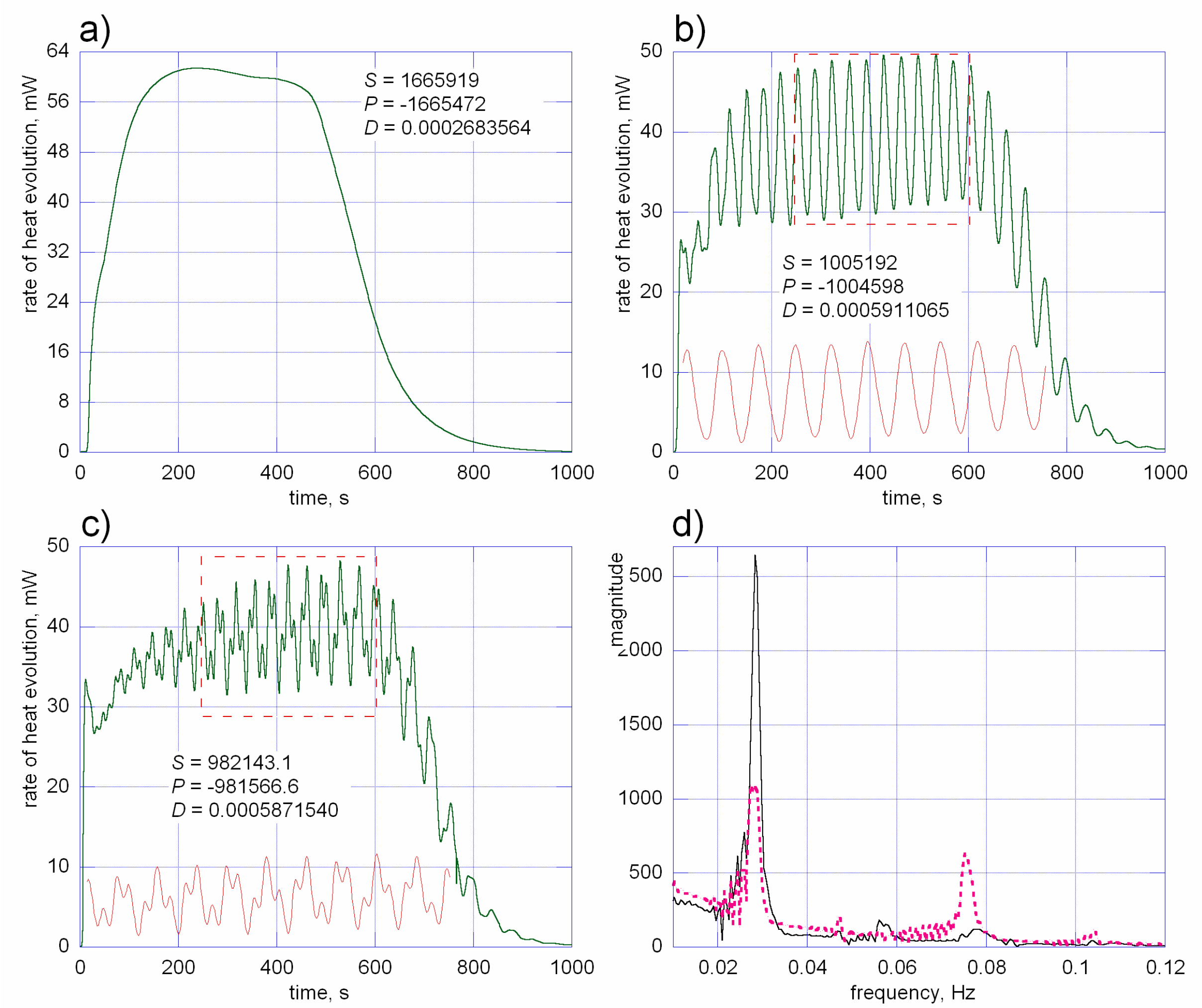

FIG. 1. Examples of non-oscillatory (**a**), periodic (**b**), and quasiperiodic (**c**) calorimetric time series representing the heat evolution in the sorption of hydrogen in metallic palladium. The corresponding power spectra for the periodic (solid black) and quasiperiodic (dashed magenta) oscillations are shown in panel (**d**). The insets (red) show expanded the sections of oscillatory curves enclosed in dashed red boxes. For explanation of *S*, *P* and *D* see text.

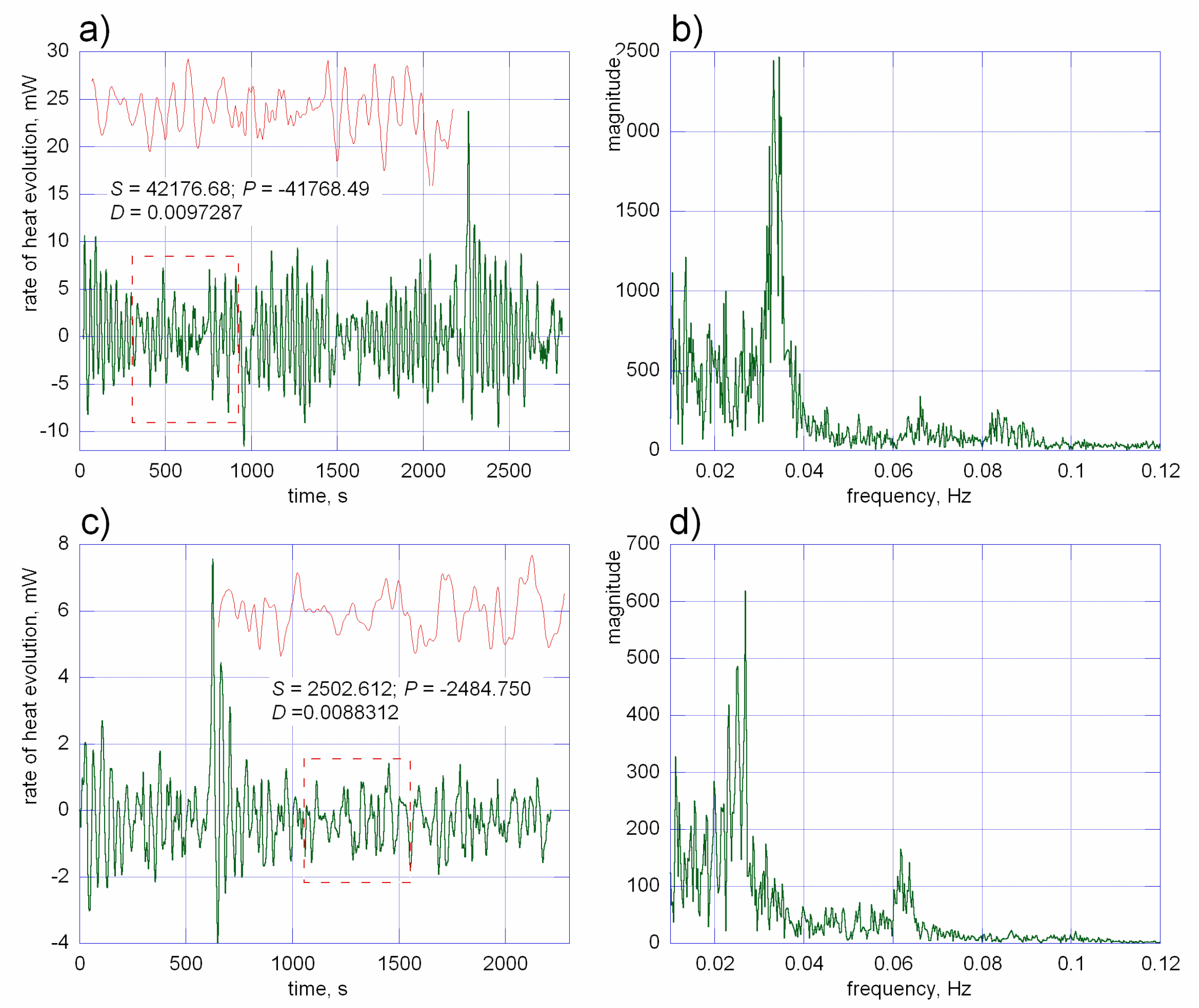

FIG. 2. Examples of aperiodic, calorimetric time series representing the heat evolution in reaction of gaseous mixtures of $H_2/N_2$ (**a**) and $D_2$/Kr (**b**) with metallic palladium at the state close to saturation, respectively, with hydrogen or deuterium. The corresponding power spectra are shown in panels (**b**) and (**d**). The insets (red) show expanded the sections of oscillatory curves enclosed in dashed (red) boxes. For explanation of *S*, *P* and *D* see text.

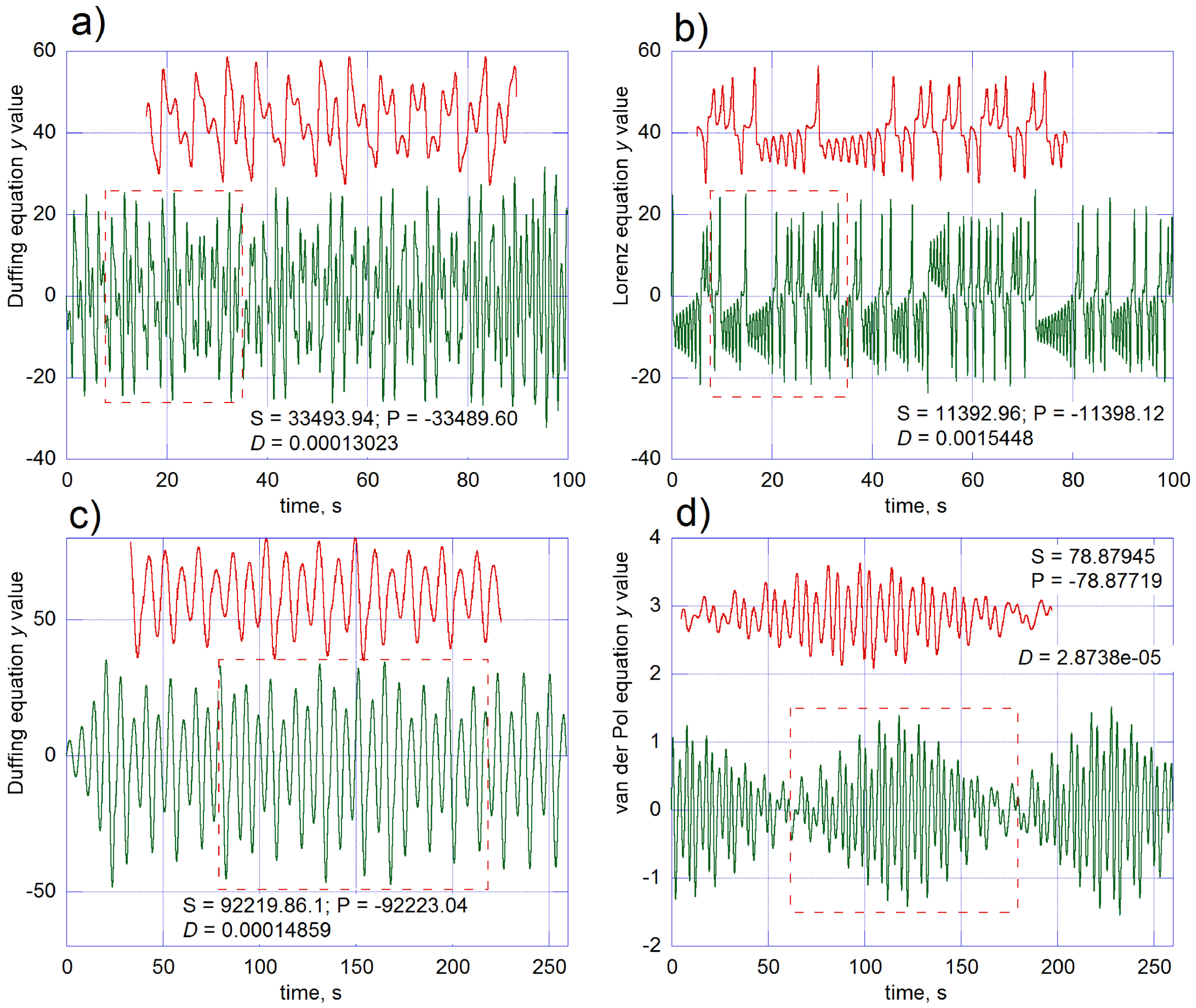

FIG. 3. Examples of chaotic time series computed using the Duffing (**a**), Lorenz (**b**), Rossler (**c**) and van der Pol (**d**) equations. The insets (red) show expanded the sections of oscillatory curves enclosed in dashed (red) boxes. For explanation of *S*, *P* and *D* see text.

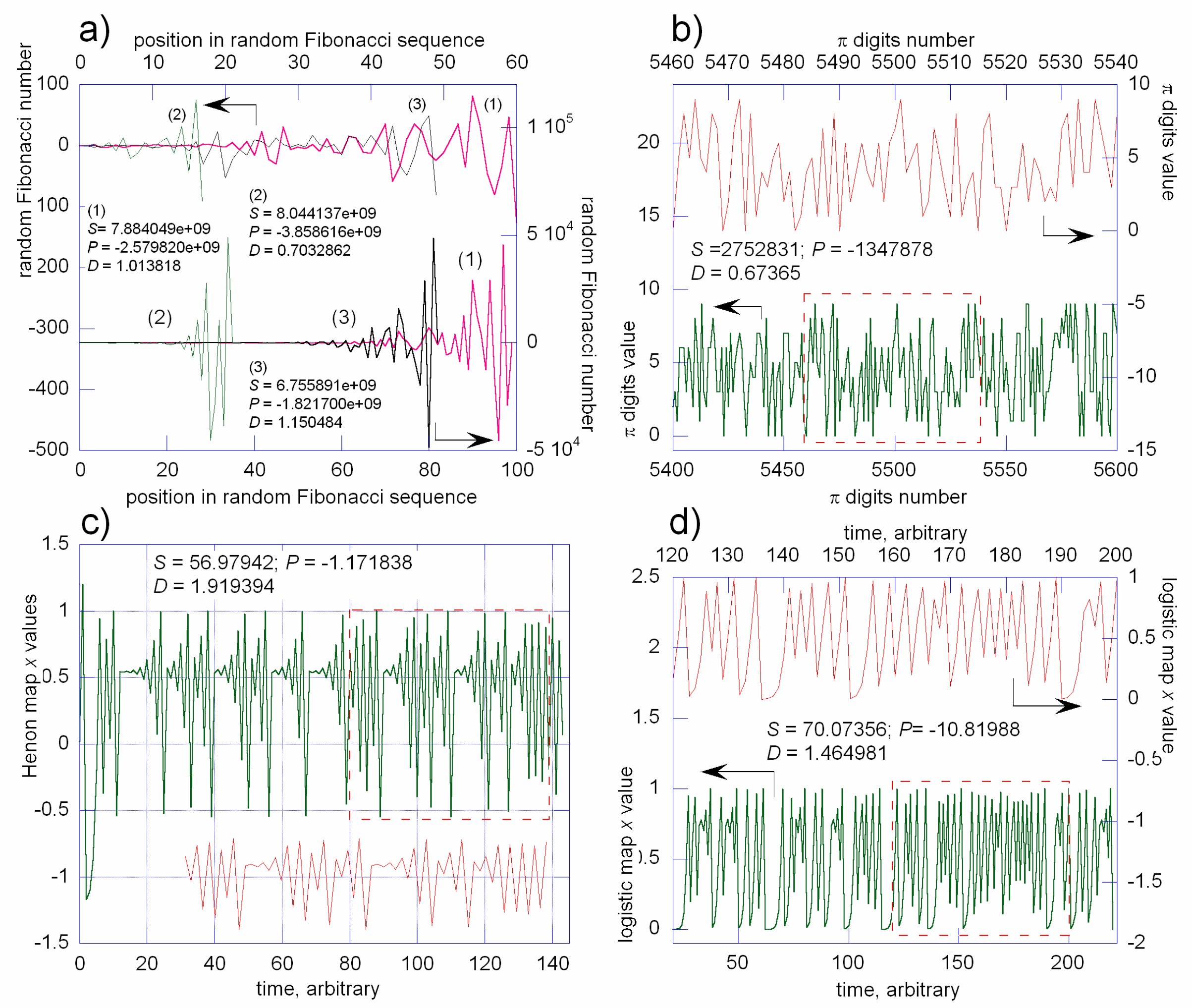

FIG. 4. Examples of the time series that do not match the conditions of Theorem 1: **a)** initial fragments (from 1 to 100) of three cases of random Fibonacci sequences, their expanded 0-to-60 regions are represented in the upper part of the panel; **b)** a sequence of digits of the number π; **c)** and **d)** the deterministic, but not continuous (discrete) time series computed using, respectively, the Henon and logistic equations. The insets (red) show expanded the sections enclosed in dashed red boxes in panels (**b**) - (**d**). For explanation of *S*, *P* and *D* see text.

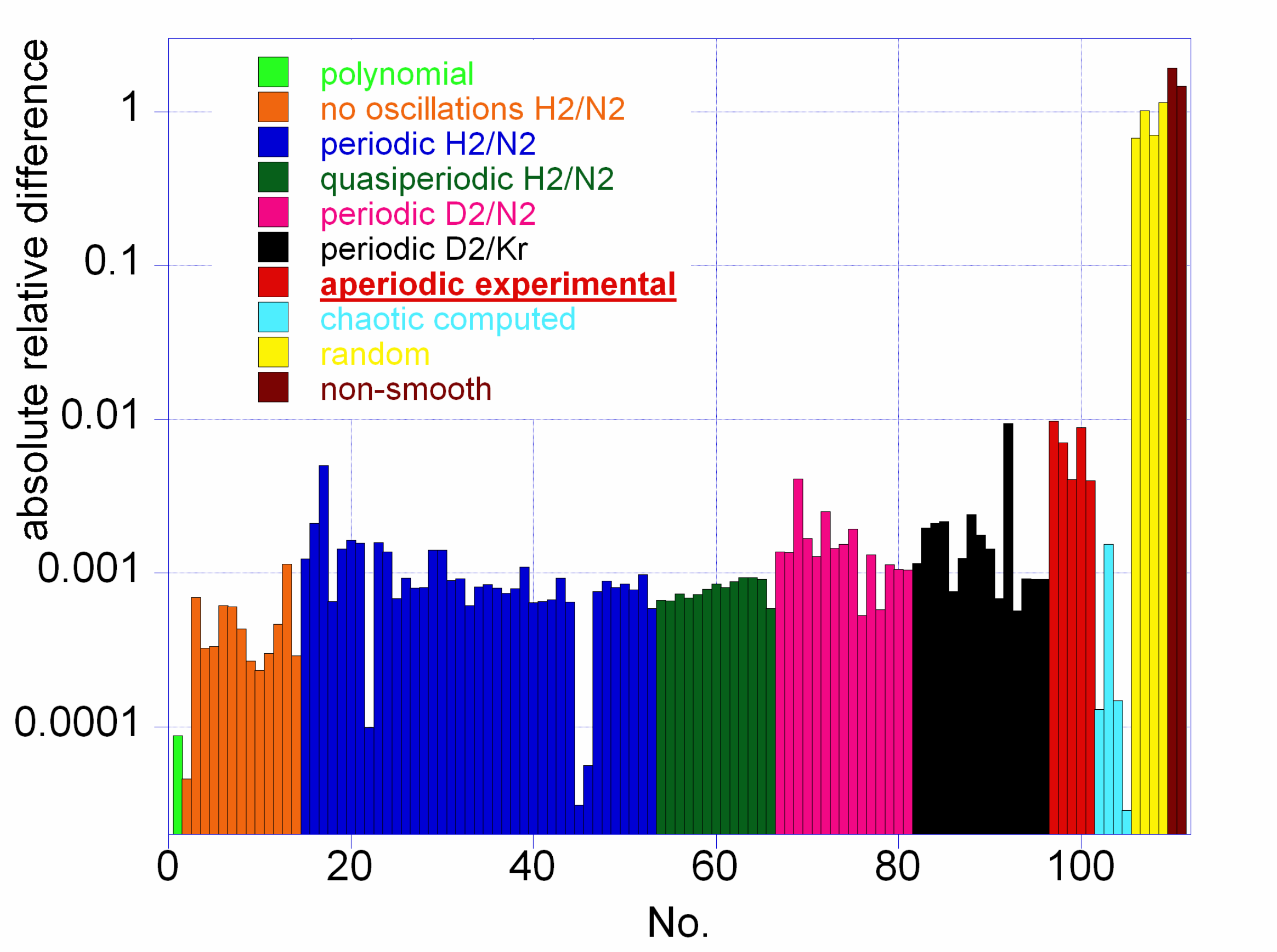

FIG. 5. Using $D$ testing for detecting determinism in aperiodic, thermokinetic oscillations recorded in the sorption of $H_2(D_2)$ in Pd. The $D$ scores obtained for the aperiodic datasets (red bars) are slightly higher than those obtained for the deterministic heat evolutions in the H(D)/Pd system, i.e., periodic oscillations (blue, magenta, black), quasiperiodic oscillations (dark green) and non oscillatory sorption (orange). However, the red bars are still two orders of magnitude lower than the $D$ scores for the random (yellow bars) and discrete (brown bars) time series (note the logarithmic scale). Numerical data are listed in Table I[10].

Supplemental material

**APPENDIX A. Differential and difference equations used to obtain time series as their numerical solutions.**

All differential equations were integrated numerically using the fourth order Runge-Kutta method for the same set of initial conditions: t = 0; $x_0$ = 5; $y_0$ = 0; $z_0$ = 5, and a step h = .003 s. A Quickbasic program for Runge-Kutta integration has been taken from Ref. 9. The program was slightly modified by the present author to enable writing of integration results into a datafile. Their chaotic dynamics has been confirmed using the power spectra and phase portraits (not shown).

The forced Duffing equation:

$$\frac{d^2x}{dt^2} + k\frac{dx}{dt} + x^3 = A\cos\Omega t \tag{A1}$$

with $k = 0.015$, $A = 68$, $\Omega = 2.5$ was recast for the Runge-Kutta integration as:

$$\begin{aligned} \frac{dx}{dt} &= y, \\ \frac{dy}{dt} &= -0.015y - x^3 + 68\cos(2.5t), \\ \frac{dz}{dt} &= 1 \end{aligned} \tag{A2}$$

The Lorenz equations:

$$\begin{aligned} \frac{dx}{dt} &= 10(y - x), \\ \frac{dy}{dt} &= x(r - z) - y, \\ \frac{dz}{dt} &= xy - \frac{8}{3}z \end{aligned} \tag{A3}$$

was integrated for r=28.

The Rossler equations:

$$\begin{aligned} \frac{dx}{dt} &= -x - z, \\ \frac{dy}{dt} &= x + 0.2y, \\ \frac{dz}{dt} &= 0.2 + (x - c)z \end{aligned} \tag{A4}$$

was integrated for c = 26.85.

The forced van der Pol equation:

$$\frac{d^2x}{dt^2} + \varepsilon(x^2 - 1)\frac{dx}{dt} + x = F\cos(\frac{2\pi}{T}t) \tag{A5}$$

with $\varepsilon = 1.8$, $F = 1.2$ and $T$=10 was recast as:

$$\begin{aligned} \frac{dx}{dt} &= y, \\ \frac{dx}{dt} &= -x - \varepsilon(x^2 - 1)y + 1.2\cos(\frac{2\pi}{10}t), \\ \frac{dz}{dt} &= 1 \end{aligned} \tag{A6}$$

The difference equations were iterated using a simple Quickbasic program written for this purpose by the author. The Henon equations:

$$\begin{aligned} x_{n+1} &= 1 - ax_n^2 + y_n, \\ y_{n+1} &= by_n \end{aligned} \tag{A7}$$

was iterated for $a = 1.55$ and $b = 0.3$ starting with $x_0 = 0.02$ and $y_0 = 0.2$.

The difference logistic equation:

$$x_{n+1} = ax_n(1 - x_n) \quad \text{(A8)}$$

was iterated with $a = 4$, and for $x_0 = 1.0\text{e-}16$.

**APPENDIX B.**

**Table I**. The values of definite integrals *S* and *P* as well as the *D*-values obtained for all the datasets tested, experimental or computed, listed in the same order as they are represented in Fig. 5. For explanation of *S*, *P* and *D* see text.

| No | Origin of dataset[a] | Dynamic character | *S* | *P* | *D* |
|---|---|---|---|---|---|
| 1 | computed, polynomial | no oscillations | 1381.209 | -1381.085 | 0.0000879 |
| 2 | experimental, $H_2/N_2$ | no oscillations | 1279613 | -1279554 | 4.610875e-05 |
| 3 | experimental, $H_2/N_2$ | no oscillations | 490092.8 | -489752.5 | 0.0006946250 |
| 4 | experimental, $H_2/N_2$ | no oscillations | 1359820 | -1359380 | 0.0003236246 |
| 5 | experimental, $H_2/N_2$ | no oscillations | 1348696 | -1348245 | 0.0003344530 |
| 6 | experimental, $H_2/N_2$ | no oscillations | 721003.8 | -720558.4 | 0.0006179928 |
| 7 | experimental, $H_2/N_2$ | no oscillations | 747933.9 | -747482.5 | 0.0006036780 |
| 8 | experimental, $H_2/N_2$ | no oscillations | 1230340 | -1229804 | 0.0004357469 |
| 9 | experimental, $H_2/N_2$ | no oscillations | 1665919 | -1665472 | 0.0002683564 |
| 10 | experimental, $H_2/N_2$ | no oscillations | 1981106 | -1980643 | 0.0002337351 |
| 11 | experimental, $H_2/N_2$ | no oscillations | 1523306 | -1522849 | 0.0003000504 |
| 12 | experimental, $H_2/N_2$ | no oscillations | 944734.7 | -944293.2 | 0.0004674362 |
| 13 | experimental, $H_2/N_2$ | no oscillations | 361325.8 | -360912.1 | 0.001145658 |
| 14 | experimental, $H_2/N_2$ | v. weak oscillations | 1580190 | -1579730 | 0.0002911466 |
| 15 | experimental, $H_2/N_2$ | periodic | 607920.5 | -607168.5 | 0.001237769 |
| 16 | experimental, $H_2/N_2$ | periodic | 192848.4 | -192441.4 | 0.002112695 |
| 17 | experimental, $H_2/N_2$ | periodic | 258586.0 | -257288.4 | 0.005030658 |
| 18 | experimental, $H_2/N_2$ | periodic | 278880.0 | -278698.5 | 0.0006510294 |
| 19 | experimental, $H_2/N_2$ | periodic | 484196.8 | -483498.4 | 0.001443443 |
| 20 | experimental, $H_2/N_2$ | periodic | 543553.3 | -542665.4 | 0.001634915 |
| 21 | experimental, $H_2/N_2$ | periodic | 523685.8 | -522868.2 | 0.001562509 |
| 22 | experimental, $H_2/N_2$ | periodic | 526565.9 | -526513.9 | 9.875796e-05 |
| 23 | experimental, $H_2/N_2$ | periodic | 806198.3 | -804926.0 | 0.001579409 |
| 24 | experimental, $H_2/N_2$ | periodic | 795080.2 | -793982.7 | 0.001381317 |
| 25 | experimental, $H_2/N_2$ | periodic | 762151.0 | -761632.3 | 0.0006807891 |
| 26 | experimental, $H_2/N_2$ | periodic | 926620.4 | -925757.9 | 0.0009312353 |
| 27 | experimental, $H_2/N_2$ | periodic | 1018720 | -1017904 | 0.0008013261 |
| 28 | experimental, $H_2/N_2$ | periodic | 956041.4 | -955270.0 | 0.0008071683 |
| 29 | experimental, $H_2/N_2$ | periodic | 1108714 | -1107155 | 0.001407123 |
| 30 | experimental, $H_2/N_2$ | periodic | 1090912 | -1089368 | 0.001416332 |
| 31 | experimental, $H_2/N_2$ | periodic | 851642.5 | -850876.6 | 0.0008996962 |
| 32 | experimental, $H_2/N_2$ | periodic | 927700.6 | -926846.9 | 0.0009207097 |
| 33 | experimental, $H_2/N_2$ | periodic | 956234.1 | -955644.2 | 0.0006171287 |
| 34 | experimental, $H_2/N_2$ | periodic | 914384.2 | -913642.4 | 0.0008115993 |
| 35 | experimental, $H_2/N_2$ | periodic | 878654.5 | -877911.1 | 0.0008463959 |
| 36 | experimental, $H_2/N_2$ | periodic | 850624.0 | -849945.9 | 0.0007975268 |
| 37 | experimental, $H_2/N_2$ | periodic | 896633.1 | -895968.9 | 0.0007411015 |
| 38 | experimental, $H_2/N_2$ | periodic | 877097.4 | -876401.6 | 0.0007935562 |
| 39 | experimental, $H_2/N_2$ | periodic | 731032.8 | -730231.0 | 0.001097423 |
| 40 | experimental, $H_2/N_2$ | periodic | 582757.0 | -582382.0 | 0.0006437000 |
| 41 | experimental, $H_2/N_2$ | periodic | 606539.1 | -606140.8 | 0.0006569128 |
| 42 | experimental, $H_2/N_2$ | periodic | 526072.4 | -525720.1 | 0.0006698089 |
| 43 | experimental, $H_2/N_2$ | periodic | 1135387 | -1134333 | 0.0009287489 |
| 44 | experimental, $H_2/N_2$ | periodic | 993104.2 | -992461.9 | 0.0006469818 |
| 45 | experimental, $H_2/N_2$ | periodic | 1095387 | -1095421 | 3.103878e-05 |
| 46 | experimental, $H_2/N_2$ | periodic | 1067982 | -1068042 | 5.617915e-05 |
| 47 | experimental, $H_2/N_2$ | periodic | 983958.3 | -983214.0 | 0.0007567334 |
| 48 | experimental, $H_2/N_2$ | periodic | 1078040 | -1077083 | 0.0008881164 |
| 49 | experimental, $H_2/N_2$ | periodic | 1083094 | -1082218 | 0.0008091213 |
| 50 | experimental, $H_2/N_2$ | periodic | 1193396 | -1192385 | 0.0008475212 |
| 51 | experimental, $H_2/N_2$ | periodic | 1194528 | -1193600 | 0.0007771777 |
| 52 | experimental, $H_2/N_2$ | periodic | 1071308 | -1070264 | 0.0009749847 |
| 53 | experimental, $H_2/N_2$ | periodic | 1005192 | -1004598 | 0.0005911065 |
| 54 | experimental, $H_2/N_2$ | quasiperiodic | 810916.0 | -810376.0 | 0.0006661354 |
| 55 | experimental, $H_2/N_2$ | quasiperiodic | 883631.1 | -883045.9 | 0.0006625433 |
| 56 | experimental, $H_2/N_2$ | quasiperiodic | 845086.9 | -844469.2 | 0.0007311832 |
| 57 | experimental, $H_2/N_2$ | quasiperiodic | 952427.7 | -951771.6 | 0.0006890692 |
| 58 | experimental, $H_2/N_2$ | quasiperiodic | 925881.9 | -925209.1 | 0.0007268686 |
| 59 | experimental, $H_2/N_2$ | quasiperiodic | 861606.0 | -860930.9 | 0.0007838729 |
| 60 | experimental, $H_2/N_2$ | quasiperiodic | 873918.0 | -873171.5 | 0.0008545641 |
| 61 | experimental, $H_2/N_2$ | quasiperiodic | 810615.1 | -809963.1 | 0.0008046511 |
| 62 | experimental, $H_2/N_2$ | quasiperiodic | 873684.9 | -872918.3 | 0.0008777752 |
| 63 | experimental, $H_2/N_2$ | quasiperiodic | 930049.4 | -929182.3 | 0.0009327105 |
| 64 | experimental, $H_2/N_2$ | quasiperiodic | 1077162 | -1076153 | 0.0009371597 |
| 65 | experimental, $H_2/N_2$ | quasiperiodic | 1076384 | -1075400 | 0.0009145900 |
| 66 | experimental, $H_2/N_2$ | quasiperiodic | 982143.1 | -981566.6 | 0.0005871540 |
| 67 | experimental, $D_2/N_2$ | periodic | 412428.5 | -411861.4 | 0.0013760 |
| 68 | experimental, $D_2/N_2$ | periodic | 444454.9 | -443851.8 | 0.0013579 |
| 69 | experimental, $D_2/N_2$ | periodic | 68121.60 | -67842.07 | 0.0041119 |
| 70 | experimental, $D_2/N_2$ | periodic | 451911.7 | -451153.3 | 0.0016796 |
| 71 | experimental, $D_2/N_2$ | periodic | 433127.9 | -432571.1 | 0.0012864 |
| 72 | experimental, $D_2/N_2$ | periodic | 527860.7 | -526528.4 | 0.0025272 |
| 73 | experimental, $D_2/N_2$ | periodic | 604171.8 | -603299.4 | 0.0014451 |
| 74 | experimental, $D_2/N_2$ | periodic | 658746.9 | -657736.6 | 0.0015348 |
| 75 | experimental, $D_2/N_2$ | periodic | 955520.6 | -953684.7 | 0.0019232 |
| 76 | experimental, $D_2/N_2$ | periodic | 708170.5 | -707795.8 | 0.00052923 |
| 77 | experimental, $D_2/N_2$ | periodic | 1085899 | -1084466 | 0.0013205 |
| 78 | experimental, $D_2/N_2$ | periodic | 767871.4 | -767425.6 | 0.00058067 |
| 79 | experimental, $D_2/N_2$ | periodic | 1192221 | -1190868 | 0.0011355 |
| 80 | experimental, $D_2/N_2$ | periodic | 1146357 | -1145143 | 0.0010596 |
| 81 | experimental, $D_2/N_2$ | periodic | 1829068 | -1827152 | 0.0010481 |
| 82 | experimental, $D_2/Kr$ | periodic | 625740.1 | -625014.8 | 0.0011598 |
| 83 | experimental, $D_2/Kr$ | periodic | 566081.6 | -564968.3 | 0.0019686 |
| 84 | experimental, $D_2/Kr$ | periodic | 735812.0 | -734266.6 | 0.0021024 |
| 85 | experimental, $D_2/Kr$ | periodic | 936012.3 | -933981.1 | 0.0021724 |
| 86 | experimental, $D_2/Kr$ | periodic | 489460.6 | -489090.8 | 0.00075577 |
| 87 | experimental, $D_2/Kr$ | periodic | 927126.2 | -925970.6 | 0.0012472 |
| 88 | experimental, $D_2/Kr$ | periodic | 923319.0 | -921106.7 | 0.0023989 |
| 89 | experimental, $D_2/Kr$ | periodic | 1200591 | -1198472 | 0.0017665 |
| 90 | experimental, $D_2/Kr$ | periodic | 1411595 | -1409570 | 0.0014356 |
| 91 | experimental, $D_2/Kr$ | periodic | 943976.0 | -943333.3 | 0.00068106 |
| 92 | experimental, $D_2/Kr$ | periodic | 1516792 | -1531095 | 0.0093855 |
| 93 | experimental, $D_2/Kr$ | periodic | 891179.7 | -890673.2 | 0.00056851 |
| 94 | experimental, $D_2/Kr$ | periodic | 1550763 | -1549340 | 0.00091803 |
| 95 | experimental, $D_2/Kr$ | periodic | 1559393 | -1557972 | 0.00091167 |
| 96 | experimental, $D_2/Kr$ | periodic | 2664409 | -2661989 | 0.00090868 |
| 97 | experimental, $H_2/N_2$ | aperiodic | 42176.68 | -41768.49 | 0.0097287 |
| 98 | experimental, $D_2/N_2$ | aperiodic | 1457.754 | -1449.471 | 0.0070473 |
| 99 | experimental, $D_2/N_2$ | aperiodic | 9066.480 | -9042.088 | 0.0040633 |
| 100 | experimental, $D_2/Kr$ | aperiodic | 2502.612 | -2484.750 | 0.0088312 |
| 101 | experimental, $D_2/Kr$ | aperiodic | 4388.287 | -4379.731 | 0.0039868 |
| 102 | computed, Duffing | chaotic | 33493.94 | -33489.60 | 0.0001302252 |
| 103 | computed, Lorenz | chaotic | 11392.96 | -11398.12 | 0.001544800 |
| 104 | computed, Rossler | chaotic | 92219.86 | -92223.04 | 0.0001485882 |
| 105 | computed, van der Pol | chaotic | 78.87945 | -78.87719 | 2.873850e-05 |
| 106 | computed, Pi n | random | 2801940 | 1352620 | 0.6736500 |
| 107 | computed, RNDFIB_1 | random | 7.884049e+09 | -2.579820e+09 | 1.013818 |
| 108 | computed, RNDFIB_2 | random | 8.044137e+09 | -3.858616e+09 | 0.7032862 |
| 109 | computed, RNDFIB_3 | random | 6.755891e+09 | -1.821700e+09 | 1.150484 |
| 110 | computed, Henon | chaotic discrete | 56.97942 | -1.171838 | 1.919394 |
| 111 | computed, logistic | chaotic discrete | 70.07356 | -10.81988 | 1.464981 |

[a] RNDFIB = random Fibonacci; three different random Fibonacci sequences have been used (denoted _1 - _3). The symbol “Pi n” denotes sequences of n digits of number $\pi$ (here the actual $D$ is an average from several sequences of the $\pi$ digits). The terms $H_2/N_2$, $D_2/N_2$ or $D_2/Kr$ (for experimental data) state the elemental composition of a gaseous mixture of hydrogen or deuterium used for reaction with metallic palladium powder.